# Warming trend in cold season of the Yangtze River Delta and its correlation with Siberian high

Chunmeng Wei[1], Jianhua Xu[2]

[1] Datong High School, Shanghai 2000011, China

[2] The Research Center for East-West Cooperation in China, School of Geographic Sciences, East China Normal University, Shanghai 200241, China1

Corresponding author: Jianhua Xu, e-mail: jhxu@geo.ecnu.edu.cn

**Abstract** Based on the meteorological data from 1960 to 2010, we investigated the temperature variation in the Yangtze River Delta (YRD) by using Mann-Kendall nonparametric test and explored the correlation between the temperature in the cold season and the Siberian high intensity (SHI) by using correlation analysis method. The main results are that (a) the temperature in YRD increased remarkably during the study period, (b) the warming trend in the cold season made the higher contribution to annual warming, and (c) there was a significant negative correlation between the temperature in the cold season and the SHI.

## 1. Introduction

With an unprecedented speed, the global warming is exacerbated by the aggravation of the greenhouse effect which induced by intensive human activities. As presented by the Fourth Assessment Report of IPCC, the global mean temperature increased 0.74℃ from 1906 to 2005, and the amount of carbon dioxide in the atmosphere was higher than ever before [1].The intensity or the frequency of extreme weather events would be aggravated due to the acceleration of climate warming [2], which have profound impacts on both human society and the natural environment. Large losses of life as well as a tremendous increase in economic losses were caused around the world, particularly in economically developed areas [3].

The Yangtze River Delta (YRD) is one of the most economically developed areas in China. This region covers 1% of China's territory area and 7.6% of its population merely. Whereas it creates 17.6% of the country's GDP, 37.6% of its exporting, and attracts 41.8% of transnational investment in 2010. The YRD was exposed to extreme weather due to the acceleration of climate warming. This region has suffered the enormous economic losses [4]. Under the background of global climate warming, research on temperature change in YRD is of great significance in sustainable development of society and economy.

A variety of studies on temperature variability in YRD were conducted in various scientific studies, including climatological [5], hydrological [6], phonological [7], and sociological [8]. The spatial patterns of the surface air temperature in YRD had six major warming centers, and it was significantly differed from the typical latitudinal pattern due to urbanization and accelerating human activities [9]. Influenced by global warming and intense urban heat islands (UHI) effects, the pronounced uptrends of temperature in YRD were detected by investigating annual mean temperature [10] and daily mean temperature [11]. The accelerating human activities and rapid urbanization is one of the significant factors for the remarkable warming trend of YRD, as considered by previous researchers [9, 11]. Higher population density, high-intensity urban construction and greenhouse gas emissions which contribute to surface warming, create and exacerbate the UHI effect in the urban areas [7, 8, 10].

Many studies emphasized accelerating human activities and rapid urbanization as the principal factors contributing to the warming trend in YRD [8-12]. Moreover, major atmospheric circulations are close related to the temperature variation as well. Song and Xu [13] discovered NAO (North Atlantic Oscillation), AO (Arctic Oscillation) and AAO (Antarctic Oscillation) have closer association with the annual mean temperature in YRD. However, the association from the Siberian high intensity (SHI) which controls the Siberia and most of China is less mentioned.

As the Siberian high is centered in northeastern Siberia [14] and its geographical location is far away from YRD, whether there is a significant correlation between the SHI and the temperature in the cold season of YRD? Whether the variability of the SHI may give some implication to the warming trend of YRD? To date, these questions have not received satisfactory answers.

Therefore, this study applies the meteorological data to examine the temperature variation in the YRD during the period of 1960-2010 and reveal the correlation between the temperature in the cold season and the SHI.

## 2. Materials and methodology
### 2.1. Study area

The Yangtze River Delta (YRD), located in the East China, is within the range of N26°96′–34°68′ and E117°17′–123°36′ (Fig. 1). The northern and middle regions of the delta are characterized by a low and flat terrain. The southern region is characterized by mountains covered with forest [5]. This typical floodplain has an average elevation of less than 10 m, and is covered with numerous rivers, lakes and channels [15]. This region is dominated by sub-tropical humid monsoon climate, with an average annual temperature from 14 to 18℃ and an annual average precipitation of approximately 1,000–1,400 mm [12]. Under the control of the East Asian monsoon, the region is occupied by a subtropical high-pressure system in the summer and under the influence of the Siberian high-pressure system in winter [16]. Multiple natural hazards, such as floods, typhoons and heat harm, pose a serious threat to the region with these geographic features.

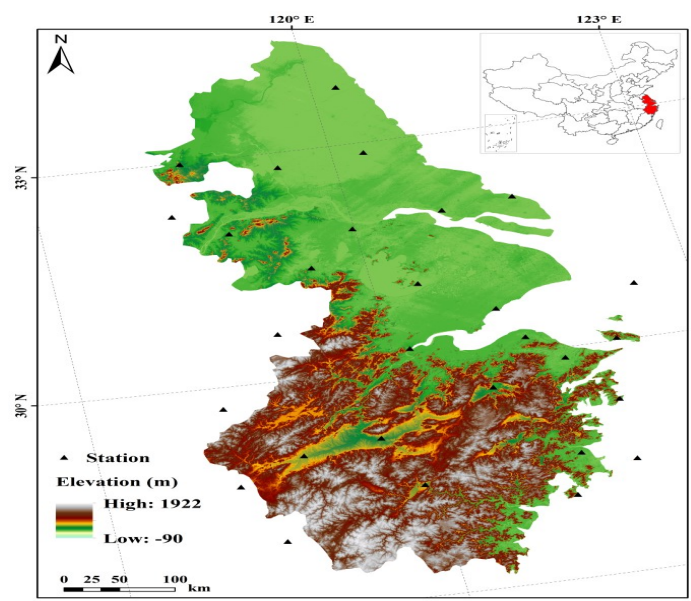

Fig. 1. The location of study area and the meteorological stations

The YRD is one of the six mega-city regions of the world. This region comprises the triangle-shaped territory of the southern Jiangsu Province, northern Zhejiang Province and Shanghai [11]. It covers an area of approximately 179,000 km$^2$ and has a population more than 81 million people in 2010. As the rapid urbanization and economic development in YRD, this region would suffer the enormous economic losses if the delta is exposed to natural hazards. That is why it is selected as the study area.

*2.2. Data sources and preprocessing*

The time series of temperature data, measured at 29 meteorological stations in the YRD during the period of 1960–2010, were obtained from the China Meteorological Data Sharing Service System (http://cdc.cma.gov.cn/). The geographical locations of meteorological stations were depicted in Fig. 1. There were 22 meteorological stations covered the whole study area. However, for more reliable interpolation results of spatial interpolation, the other 7 stations, being located around the YRD, were selected to analyze the temporal and spatial variation of temperature as well.

The SHI index was used to analyze the variability of high and the possible impact on global warming by Gong and Wang [17]. SHI was calculated as the regional mean sea level pressure averaged over the 60°E -130°E and 30°N -70°N where the pressure was greater than 1028 hPa [18, 19]. Finally, we obtained the SHI based on the 1960–2010 NCEP (National Centers for Environmental Prediction)/ NCAR (National Center for Atmospheric Research) reanalysis.

The Siberian high forms generally in November and persists until disappears around the end of April [17, 19, 20, 21, 42]. Accordingly, we calculated the temperature trends for the cold season and warm season, with the cold season as defined from November to April of next year, and warm season from May to October. In order to detect the temporal trends of temperature in YRD, we calculated the annual, cold and warm season mean temperature values by using the arithmetic mean series of all meteorological stations.

There were four months with missing temperature data, in August, September and October 1967 for Gaoyou station, in August 1967 for Tunxi station. To make the data integrity, we adopted ratio method [22] to deal with the missing data in this paper.

*2.3. Methodology*

(1) Trend Test

The Mann-Kendall (MK) trend test [23, 24], one of the non-parametric statistical tests for detecting trends in the time series, has been frequently applied in hydro-meteorological time series [25- 28]. Compared with parametric statistical test, the non-parametric test is more suitable for non-normally distributed data [26].

At a 5 % significance level, the null hypothesis $H_0$ for the Mann-Kendall test which sample of data $X=\{x_1, x_2…, x_n\}$ has no trend, is rejected if $|Z| > 1.96$. The alternative hypothesis $H_1$ is that a monotonic trend exists in $X$. A positive value of Z denotes a rising trend, and the opposite presents a decreasing trend. When $n \geq 8$, the standard normal statistic Z is computed as following:

$$Z = \begin{cases} \dfrac{S-1}{\sqrt{Var(S)}}, & S > 0 \\ 0, & S = 0 \\ \dfrac{S+1}{\sqrt{Var(S)}}, & S < 0 \end{cases} \quad (1)$$

in which

$$S = \sum_{i=1}^{n-1} \sum_{j=i+1}^{n} \operatorname{sgn}(x_j - x_i),$$

$$\operatorname{sgn}(\theta) = \begin{cases} 1, & \theta > 0 \\ 0, & \theta = 0 \\ -1, & \theta < 0 \end{cases}$$

$$Var(S) = \frac{n(n-1)(2n+5) - \sum_{i=1}^{n} t_i i(i-1)(2i+5)}{18}$$

where the $x_j$, $x_i$ are the sequential data values, $n$ is the length of the data set, and $t_i$ is the number of ties of extent $i$.

(2) Correlation analysis

Correlation analysis is a statistical measurement of the systematic relationship between two variables [29]. For variables x and y, such as the SHI and temperature, the Pearson's correlation coefficient is calculated as follows:

$$r_{xy} = \frac{\sum_{j=1}^{n}(x_i - \bar{x})(y_i - \bar{y})}{\sqrt{\sum_{i=1}^{n}(x_i - \bar{x})^2}\sqrt{\sum_{i=1}^{n}(y_i - \bar{y})^2}} \qquad (2)$$

Where $x_i$ and $y_i$ represent the values of $x$ and the values of $y$ for the sample $i$; $\bar{x}$ and $\bar{y}$ are the means for all $x_i$ and all $y_i$; $n$ is the sample number. And $t$ distribution is commonly employed to test for the significance of the correlation coefficient.

## 3. Results and discussions

### 3.1. Warming trend in YRD

For checking up the long-term trend of temperature in YRD over the study period, the arithmetic means temperature series of all meteorological stations were analyzed at inter-annual, cold season and warm season scales (Fig. 2). Investigating on different time scales help identify the contribution values of seasonal temperature change in the yearly change.

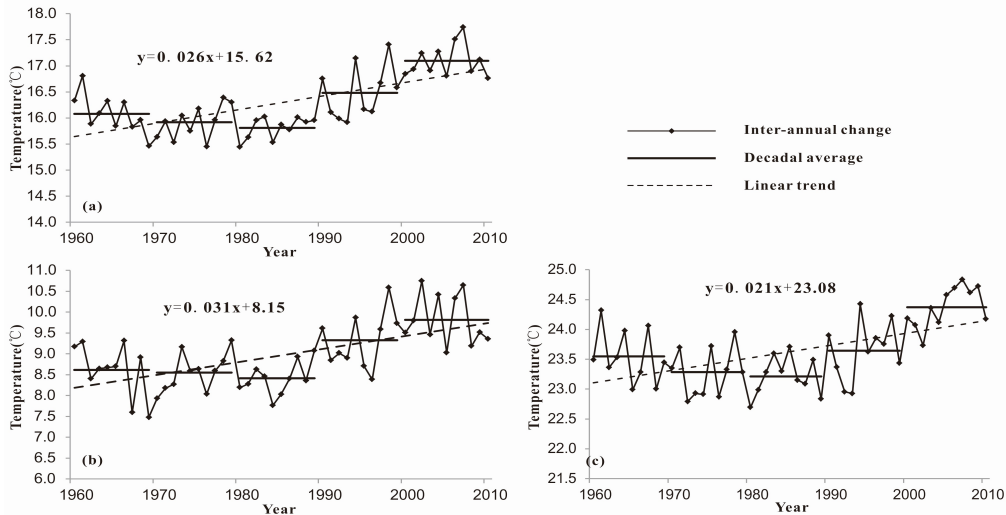

Fig. 2. The trends of temperature in the YRD during the period of 1960-2010: (a) annual, (b) cold season, (c) warm season

The change trends of the annual and seasonal (the cold season and the warm season) mean temperature were the same, a rising trend, which exhibited an obvious inter-decadal shift during the period of 1960-2010. For the annual mean temperature, it was fairly cold from the early 1960s to the 1980s, and the pronounced warming trend began in the late 1980s. Since then, the annual average temperature has increased rapidly. With regards to the seasonal mean temperature, a similar pattern, gradual decline to cold and a sudden ascending after then (warn-cold-warm), was also noted in cold and warm season. The annual, cold and warm season mean temperature rose sharply to a peak in 2000s, which were more than 1.28℃, 1.41℃ and1.16℃ respectively, as compared with minimum inter-decadal mean temperature.

To provide results which were statistically significant, we test all linear trends of the annual, cold season and the warm season mean temperature by applying nonparametric Mann–Kendall trend test (Table 1).

Table 1. Change trends, variability and its contribution of annual and seasonal mean temperature

|  | Annual | Cold season | Warm season |
| --- | --- | --- | --- |
| Mean Value (℃/a) | 16.29 | 8.96 | 23.63 |
| Z values | 4.32 | 4.16 | 3.62 |
| Temperature variability (℃/10a) | 0.26 | 0.31 | 0.21 |
| Contribution of variability (%) | - | 59.6 | 40.4 |

**Notes:** all the statistic Z values passed the significance level of 0.01; contribution of variability (%) indicate the percentage of the seasonal (cold and warm season) temperature variability in annual temperature variability.

The test results showed that all the statistic Z values passed the significance test of 0.01. There was a significant warming trend in the Yangtze River Delta (YRD) from 1960 to 2010. Moreover, the increasing trend was revealed by the temperature variability rates. Annual mean temperature variability rate has reached 0.26℃/10a, indicating the more obvious warming, compared to the warming rate all over china(0.22℃/10a) and that of the world(0.13℃/10a) in the past 50 years.

Nevertheless, differences were discovered between the temperature rate in cold season and that in warm season: the warming rate in the cold season is the higher than that in warm season. Warming rates for the cold season and the warm season in YRD were 0.31℃/10a and 0.21℃/10a (Table 1). This result can be confirmed by the previous studies. Just as considered by Pan et al [30], contribution of warming trend in YRD was the maximum in winter, then in spring and autumn, and the minimum in summer. And Jiang [31] also hold that the variability of warming displayed higher values in the winter and spring, while it was relatively lower in the summer and autumn, presenting the most intense warming appeared in the cold season (winter and spring). Under the background of climate warming, the warming trend of yearly temperature was dominantly attributed to the variability of temperature in the cold season, and the temperatures in the cold season were highly sensitive to annual temperature change.

In order to indicate the temperature variation trend of 29 meteorological stations in YRD, we calculated the annual and cold season mean temperature variability of each station and compared the contribution values of seasonal warming to the annual warming (Table 2). The spatial difference of the temperature variability (Fig. 3) was analyzed by the Ordinary Kriging method [32, 33].

Spatial distributions of the annual and cold season mean temperature variation (Fig. 3) displayed that the temperature for 29 meteorological stations in YRD showed the pronounced uptrend, whereas the obvious spatial variation existed in the warming change: the warming rate showed a zonal distribution. The spatial pattern of temperature variation differed from the typical latitudinal pattern which increased spatially with increasing latitude. Specifically, the warming trend was fairly fast at the lower reaches of the Yangtze River and the Hangzhou Bay, while it was rather slow in the southwest of Zhejiang province.

Table 2. Contribution values of the cold season warming to the annual warming

| Station | Warming rate(℃/10a) | | Contribution (%) to annual warming |
|---|---|---|---|
| | Annual | Cold season | Cold season |
| Xuyi | 0.20 | 0.31 | 78.4 |
| Sheyang | 0.27 | 0.35 | 64.4 |
| Chuzhou | 0.30 | 0.36 | 60.0 |
| Nanjing | 0.28 | 0.35 | 62.1 |
| Gaoyou | 0.32 | 0.39 | 60.5 |
| Dongtai | 0.22 | 0.29 | 65.0 |
| Nantong | 0.33 | 0.37 | 57.1 |
| Lvsi | 0.28 | 0.35 | 62.1 |
| Changzhou | 0.31 | 0.36 | 58.9 |
| Suyang | 0.31 | 0.35 | 55.1 |
| Dongshan | 0.23 | 0.27 | 57.7 |
| Ningguo | 0.20 | 0.24 | 62.4 |
| Hangzhou | 0.36 | 0.40 | 56.5 |
| Pinghu | 0.27 | 0.31 | 58.2 |
| Cixi | 0.36 | 0.39 | 54.5 |
| Shengsi | 0.26 | 0.32 | 61.6 |
| Dinghai | 0.20 | 0.21 | 54.2 |
| Tunxi | 0.20 | 0.22 | 55.8 |
| Jinhua | 0.24 | 0.30 | 62.9 |
| Shengzhou | 0.19 | 0.24 | 60.5 |
| Yinzhou | 0.43 | 0.46 | 53.3 |
| Shipu | 0.22 | 0.28 | 63.2 |
| Quzhou | 0.12 | 0.17 | 71.7 |
| Yushan | 0.15 | 0.22 | 69.8 |
| Lishui | 0.18 | 0.24 | 65.5 |
| Hongjia | 0.31 | 0.36 | 57.9 |
| Dachen | 0.46 | 0.50 | 54.6 |
| Yuhuan | 0.22 | 0.27 | 60.7 |
| Pucheng | 0.12 | 0.15 | 64.3 |
| Minimum | 0.12 | 0.15 | 53.3 |
| Maximum | 0.46 | 0.50 | 78.4 |

**Notes:** Minimum and Maximum are the minimum values and the maximum values of statistics for all meteorological stations.

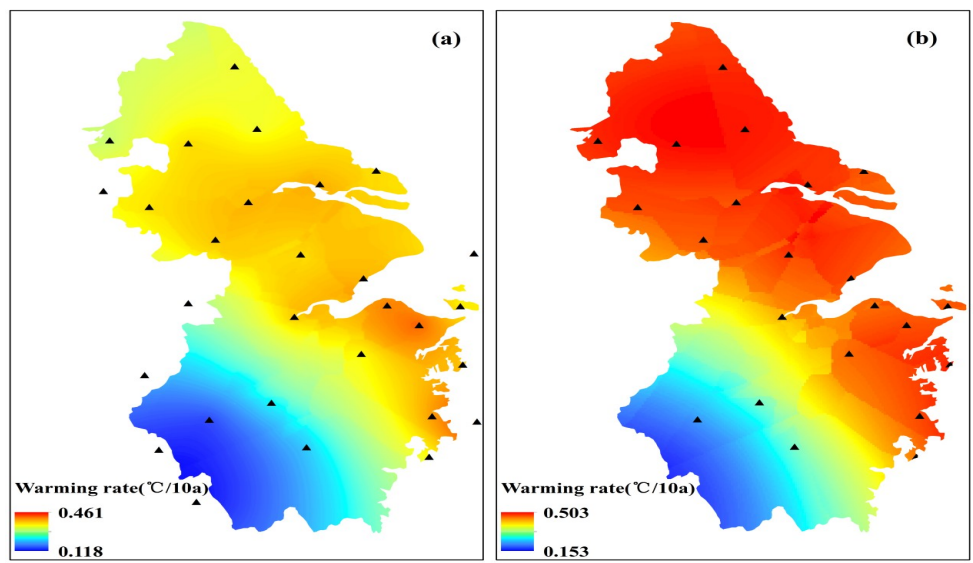

Fig. 3. Spatial distributions of (a) annual and (b) cold season warming rate in the Yangtze River Delta

Compared with the annual warming rates, cold season warming rates displayed higher values in all meteorological stations (Table 2). Consequently, the temperature in the cold season showed the more remarkable uptrend than that in the warm season. Furthermore, the contribution values of the cold season warming to the annual warming reached more than 50% in all meteorological stations, with the minimum (53.3%) in Yinzhou station and maximum (78.4%) the in Xuyi station. The temperature in the cold season was highly sensitive to annual temperature warming. These results suggest that the significant warming trend in YRD responded to regional and global warming and showed the evident of seasonal differences characteristic remarkably.

On the whole, the variability of temperature in YRD showed the increasing trend. Under the background of warming trend, the most intense seasonal warming appeared in the cold season. This conclusion is consistent with other studies, which is not only in the YRD, but also in other regions of China, such as in the Northwest China, in the Northeast and North of China, etc. [34, 35]. The emission intensity of greenhouse gases appears to increase rapidly in the cold season with the accelerating human activities, leading to the enhancement of the atmospheric radiation heat preservation, which is one of the main reasons for the higher warming rate appearing in the colder half of the year [10, 36].

### *3.2. Correlation between the temperature in the cold season and the Siberian high intensity*

It is generally known that East Asia winter monsoon, controlled by the Siberian high intensity (SHI), principally influences the weather in the cold season. The Siberian high, also called Siberian anticyclone, is the dominant atmospheric circulation system in the lower troposphere [14, 18]. The expansion of the Siberian high controls the Siberia and northern China during the colder half of the year [37, 38]. And SHI has close relation to cold air outbreak and active cold surges [39, 40]. Due to this correlation, the changes of the SHI give an explanation of temperature variability in the cold season. For example, in the arid region of northwest China, the winter temperature change has the greatest importance for the annual temperature variation. And it has a significant association with the Siberian High. Therefore, a weak Siberian High result in a relatively high winter temperature in this region [19].

During the study period, linear trend coefficients of the SHI was computed as negative value, showing the weakening of the SHI (Fig. 4). While the temperature in the cold season of the YRD showed the increasing trend. Some previous studies have shown that there were very strong couplings between the SHI and surface temperature in northern China [35, 41]. It is unclear about whether there is a significant correlation between the SHI and the temperature in the cold season of YRD.

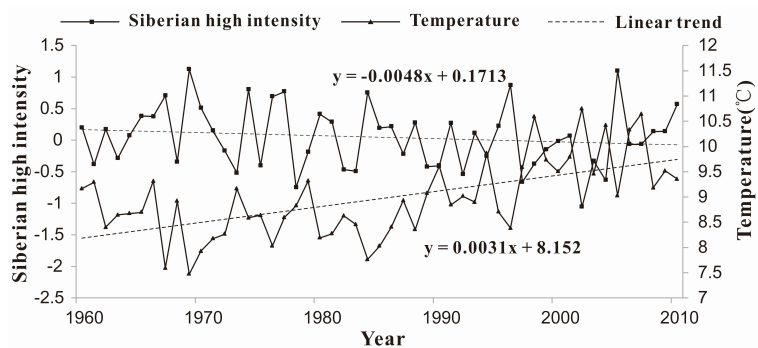

Fig. 4. Temperature in the cold season and the Siberian high intensity during the period of 1960-2010

In this study, we computed the Pearson correlation coefficient between the SHI and the temperature in the cold season of YRD, which was -0.57, indicating the negative correlation at the 1% confidence level significantly. Table 3 showed that, the temperature in cold season for all meteorological stations had significant relationship with the SHI, with the highest correlation (-0.62) in Quzhou station and the lowest correlation (-0.48) the in Hongjia station.

Table 3. The Correlation coefficients between the temperature in the cold season and the Siberian high intensity

| Station | Correlation | Station | Correlation |
| --- | --- | --- | --- |
| Xuyi | -0.571* | Shengsi | -0.554* |
| Sheyang | -0.538* | Dinghai | -0.592* |
| Chuzhou | -0.546* | Tunxi | -0.559* |
| Nanjing | -0.54* | Jinhua | -0.552* |
| Gaoyou | -0.509* | Shengzhou | -0.595* |
| Dongtai | -0.588* | Yinzhou | -0.489* |
| Nantong | -0.542* | Shipu | -0.578* |
| Lvsi | -0.519* | Quzhou | -0.62* |
| Changzhou | -0.568* | Yushan | -0.566* |
| Suyang | -0.534* | Lishui | -0.549* |
| Dongshan | -0.605* | Hongjia | -0.48* |
| Ningguo | -0.603* | Dachen | -0.503* |
| Hangzhou | -0.549* | Yuhuan | -0.569* |
| Pinghu | -0.519* | Pucheng | -0.537* |
| Cixi | -0.491* | | |
| Minimum | -0.48* | Maximum | -0.62* |

**Notes:** *correlated at significance level of 0.01.

Due to the closely correlates, the changes of the SHI give an explanation of the variations of the temperature in cold season of YRD: with the strengthened Siberian high, the winter monsoon is stronger, which correspond to stronger cold waves in the YRD, following the temperature decreasing intensely; rather, a remarkable warming tendency relates to the weakening of the SHI [18].

There are strong tendencies for the warming trend associated with the weakened SHI, which can explain the warming trends in the cold season to some extent. That is to say, from 1960 to 2010, because of the influence of the weakened SHI, the warming trends in the cold season were more significant in the YRD. In summary, the intensity of the Siberian high was closely related to the temperature in the YRD during the colder half of the year.

**4. Summary and concluding remarks**

In summary, the warming trend in the YRD in the past 50 years cannot be entirely ascribed to the accelerating human activities and rapid urbanization, while the atmospheric circulation is one of the most important induced factors, particularly the SHI. It is obvious that the temperature in YRD is influenced drastically by the SHI, principally during the cold half of the year.

Based on the results and discussions above, we analyzed the main conclusions as follows:

(1) In the past 50 years, the annual, cold and warm season mean temperature displayed an obvious uptrend in the YRD. However, the temperature variability rate possessed the seasonal differences. The warming rate in the cold season was higher than that in the warm season. Moreover, the variability of temperature in the cold season made the higher contribution than that in the warm season to annual warming. The warming trend of annual temperature was dominantly attributed to the variability of temperature in the cold season.

(2) The annual and cold season mean temperature variation in all meteorological stations showed the pronounced warming. The contribution values of the cold season warming to the annual warming reached more than 50% in all meteorological stations. Under the background of warming trend, the most intense seasonal warming in all meteorological stations of YRD appeared in the cold season.

(3) The weakening of the SHI during the past 50 years is its most striking feature. Since the temperature in cold season of the YRD was negatively correlated with the SHI at 1% significance level, the weakened SHI is one of the main reasons for the warming trends during the colder half of the year.

**Conflict of Interests**

The authors declare that there is no conflict of interests regarding the publication of this article.